\documentstyle [12pt] {article}
\begin{document}
\def\be{\begin{eqnarray}}
\def\en{\end{eqnarray}}
\def\non{\nonumber}
\def\la{\langle}
\def\ra{\rangle}
\def\ep{\varepsilon}
\def\b{\Lambda_b\to J/\psi\Lambda}
\def\j{{J/\psi}}
\def\pr{{\sl Phys. Rev.}~}
\def\prl{{\sl Phys. Rev. Lett.}~}
\def\pl{{\sl Phys. Lett.}~}
\def\np{{\sl Nucl. Phys.}~}
\def\zp{{\sl Z. Phys.}~}

\font\el=cmbx10 scaled \magstep2
{\obeylines
\hfill IP-ASTP-22-94
\hfill November, 1994}

\vskip 1.5 cm

\centerline{\large\bf The $\Lambda_b\to J/\psi+\Lambda$ Decay Revisited}
\medskip
\bigskip
\medskip
\centerline{\bf Hai-Yang Cheng and B. Tseng}
\medskip
\centerline{ Institute of Physics, Academia Sinica}
\centerline{Taipei, Taiwan 11529, Republic of China}
\bigskip
\bigskip
\bigskip
\centerline{\bf Abstract}
\bigskip
{\small
   The $\b$ decay does not receive any nonspectator contributions and hence
its theoretical calculation is relatively clean. Two different methods are
employed for the study of $\b$. In the first method, both $b$ and $s$ quarks
are treated as heavy and leading $1/m_s$ as well as $1/m_b$
corrections are included. The
branching ratio and the asymmetry parameter $\alpha$ are found to be
$(0.7-1.5)\times 10^{-3}$ and $0.18$, respectively. In the second method,
only the $b$ quark is treated as heavy. The
prediction of the decay rate is very sensitive to the choice of the
$q^2$ dependence for form factors. We obtain $\alpha=0.31$ and ${\cal B}(\b)
\simeq 1.6\times 10^{-4}$ for dipole $q^2$ behavior as well as $8.2\times
10^{-4}$ for monopole $q^2$ dependence. We conclude that ${\cal B}(\b)~{\large
^<_{\sim}}~10^{-3}$ and $\alpha$ is of order $0.20\sim 0.30$ with a positive
sign.
}

\pagebreak
   The $\b$ decay was originally reported by the
UA1 Collaboration [1] with the result
\be
F(\Lambda_b){\cal B}(\b)=\,(1.8\pm 0.6\pm 0.9)\times 10^{-3},
\en
where $F(\Lambda_b)$ is the fraction of $b$ quarks fragmenting into
$\Lambda_b$. Assuming $F(\Lambda_b)=10\%$ [1], this leads to
\be
{\cal B}(\b)=\,(1.8\pm 1.1)\%.
\en
However, both CDF [2] and LEP [3] did not see any evidence for this decay.
For example, based on the
signal claimed by UA1, CDF should have reconstructed $30\pm 23$ $\b$ events.
Instead CDF found not more than 2 events and concluded that
\be
F(\Lambda_b){\cal B}(\b)< 0.50\times 10^{-3}.
\en
The limit set by OPAL is [3]
\be
F(\Lambda_b){\cal B}(\b)< 1.1\times 10^{-3}.
\en
Hence, a theoretical estimate of the branching ratio for this decay mode
would be quite helpful to clarify the issue.

    The exclusive two-body decays of the $\Lambda_b$ is predominated by
$\Lambda_b\to\Lambda_c^{(*)}\pi(\rho)$. However, the decay $\b$ is of
particular interest from the theoretical point of view. At the quark level,
the nonleptonic weak decays of the baryon usually receive contributions
from external $W$-emission, internal $W$-emission and $W$-exchange diagrams.
At the hadronic level, these contributions manifest as factorizable and
pole diagrams. It is known that, contrary to the meson case, the nonspectator
$W$-exchange effects in charmed baryon decays are of comparable importance
as the spectator diagrams [4]. Unfortunately, in general it is difficult to
estimate the pole diagrams. Nevertheless, there exist some decay modes
of bottom baryons which proceed only through the internal or external
$W$-emission diagram. Examples are
\be
&& {\rm internal}~W{\rm -emission}:~~
\b,~~~\Xi_b\to J/\psi\Xi,~~~\Omega_b\to J/\psi\Omega, \non \\
&& {\rm external}~W{\rm -emission}:~~
\Omega_b\to\Omega_c\pi.
\en
Consequently, the $\b$ decay is free of nonspectator effects and its
theoretical calculation is relatively clean.

   The $\b$ decay has been considered in Ref.[5] using the heavy $s$
quark approach. The purpose of this short Letter is to update the analysis of
Ref.[5] and furthermore present a quark-model estimate of its decay rate.
Theoretical uncertainties are then addressed.

    The general amplitude of $\b$ has the form
\be
A(\b)=i\bar{u}_\Lambda(p_\Lambda)\varepsilon^{*\mu}[A_1\gamma_\mu\gamma_5+
A_2(p_\Lambda)_\mu\gamma_5+B_1\gamma_\mu+B_2(p_\Lambda)_\mu]u_{\Lambda_b}
(p_{\Lambda_b}).
\en
Under factorization assumption, the internal $W$-emission contribution reads
\be
A(\b)=\,{G_F\over\sqrt{2}}V_{cb}V_{cs}^*a_2\la J/\psi|\bar{c}\gamma_\mu(1-
\gamma_5)c|0\ra\la\Lambda|\bar{s}\gamma^\mu(1-\gamma_5)b|\Lambda_b\ra,
\en
where $a_2$ is an unknown parameter introduced in Ref.[6]. It follows from
Eqs.(6) and (7) that
\be
A_1 &=& -\eta[g_1(m^2_\j)+g_2(m^2_\j)(m_{\Lambda_b}-m_\Lambda)],   \non \\
A_2 &=& -2\eta g_2(m^2_\j),   \non \\
B_1 &=& \eta[f_1(m^2_\j)-f_2(m^2_\j)(m_{\Lambda_b}+m_\Lambda)],   \\
B_2 &=& 2\eta f_2(m^2_\j),   \non
\en
with $\eta={G_F\over\sqrt{2}}V_{cb}V_{cs}^*a_2f_\j m_\j$, where $f_i$
and $g_i$ are the form factors defined by
\be
\la\Lambda(p_\Lambda)|\bar{s}\gamma_\mu(1-\gamma_5)b|\Lambda_b(p_{\Lambda_b})
\ra &=& \bar{u}_\Lambda
[f_1(q^2)\gamma_\mu+if_2(q^2)\sigma_{\mu\nu}q^\nu+f_3(q^2)q_\mu   \non \\
&& -(g_1(q^2)\gamma_\mu+ig_2(q^2)\sigma_{\mu\nu}q^\nu+g_3(q^2)q_\mu)\gamma_5]
u_{\Lambda_b},
\en
where $q=p_{\Lambda_b}-p_\Lambda$.
Evidently, the main task is to compute the form factors.

  In principle, one can apply the quark model to calculate the form factors
$f_i$ and $g_i$ directly. However, the difficulty is that, as we shall see,
the value of the form factors at $q^2=m^2_\j$ is very sensitive to the choice
of their $q^2$ dependence. Instead we would like to appeal to the methods in
which model dependence is kept to minimum. We will consider two different
methods to estimate the rate of $\b$.
\footnote{These two methods were already elaborated on in Ref.[7] for the
$\Lambda_b\to\Lambda\gamma$ decay.}
In the first method, both $b$ and $s$ quarks are treated as heavy. Since the
$s$ quark mass in the baryon is only of order 500 MeV, it is important to
estimate the $1/m_s$ corrections to see if it makes sense to apply heavy
quark symmetry to the $s$ quark at first place. In the second method, only
the $b$ quark is treated as heavy. Using the recent CLEO measurement for
the form-factor ratio in $\Lambda_c-\Lambda$ transition, one can deduce the
form-factor ratio in the $\Lambda_b-\Lambda$ matrix element.

   In the heavy $b$-quark and $s$-quark limit, the form factors are simply
given by $f_1=g_1=1$ at zero recoil and $f_2=g_2=f_3=g_3=0$ [8]. Following
Ref.[9] to add the $1/m_s$ and $1/m_b$ corrections, we obtain
\footnote{The $(1-m_{\Lambda}/m_{\Lambda_b})$ term in Eq.(27) of Ref.[5]
should read $(\omega-m_{\Lambda}/m_{\Lambda_b})$. Contrary to some claims
made in the literature, it is not necessary to introduce a Clebsch-Gordon
coefficient $1/\sqrt{3}$ in Eq.(10).}
\be
&& f_1=g_1=\left[ 1+{\bar{\Lambda}\over 2m_s}\,{1\over 1+\omega}\left(\omega
-{m_\Lambda\over m_{\Lambda_b}}\right)+{\bar{\Lambda}\over 2m_b}\,{1\over
1+\omega}\left(\omega-{m_{\Lambda_b}\over m_{\Lambda}}\right)\right]h,
   \non  \\
&& f_2=g_3=-{\bar{\Lambda}\over 2(1+\omega)}\left({1\over m_sm_{\Lambda_b}}
+{1\over m_bm_\Lambda}\right)h,  \\
&& f_3=g_2=-{\bar{\Lambda}\over 2(1+\omega)}\left({1\over m_sm_{\Lambda_b}}
-{1\over m_bm_\Lambda}\right)h, \non
\en
with $\omega\equiv v_{\Lambda_b}\cdot v_{\Lambda}=1.86,~\bar{\Lambda}=m_{
\Lambda_b}-m_b=m_{\Lambda_c}-m_c=m_\Lambda-m_s\approx 700$ MeV, and
\be
h=\left({\alpha_s(m_b)\over \alpha_s(m_c)}\right)^{-6/25}
\left({\alpha_s(m_c)\over \alpha_s(m_s)}\right)^{-6/27}
\left({\alpha_s(m_s)\over \alpha_s(\mu)}\right)^{a_L(\omega)}\zeta(\omega,\mu),
\en
where $\zeta(\omega)$ is a universal baryonic Isgur-Wise function normalized
to unity at zero recoil or $v_{\Lambda_b}\cdot v_\Lambda=1$,
and the expression for $a_L(\omega)$ can be found in
Ref.[10]. As in Ref.[5] we choose the normalization scale $\mu\sim m_s$ so
that $\alpha_s(\mu)\sim 1$ and $h\simeq 1.23\,\zeta(\omega,\mu)$. We see from
Eq.(10) that the $1/m_s$ correction to the $\b$ amplitude is quite sizeable,
about 40\% for $m_s=500$ MeV. Therefore, higher order $1/m_s$ corrections
should be included for a realistic comparsion with experiment. In practice,
however, it is impossible to carry out this formidable task in terms of the
present technique.

    In order to estimate the $\b$ rate, we employ two recent models for
$\zeta(\omega)$:
\be
\zeta(\omega)=\cases{ 0.99\exp[-1.3(\omega-1)], & soliton~model~[11]; \cr
\left({2\over \omega+1}\right)^{3.5+{1.2\over\omega}}, & MIT~bag~model~[12].
\cr}
\en
Hence, $\zeta(\omega=1.86)$ ranges from 0.23 to 0.32. The decay rate
is given by [13]
\be
\Gamma(\b)={1\over 8\pi}\,{E_\j+m_\j\over m_{\Lambda_b}}\,p_c\left[
2(|S|^2+|P_2|^2)+{E^2_\j\over m_\j^2}(|S+D|^2+|P_1|^2)\right],
\en
with
\be
S &=& -A_1,   \non \\
D &=& {p_c^2\over E_\j(E_\Lambda+m_\Lambda)}\,(A_1-m_{\Lambda_b}A_2),\non\\
P_1 &=& -{p_c\over E_\j}\left({m_{\Lambda_b}+m_\Lambda\over E_\Lambda+m
_\Lambda}B_1+m_{\Lambda_b}B_2\right),    \\
P_2 &=& {p_c\over E_\Lambda+m_\Lambda}B_1,   \non
\en
where $p_c$ is the c.m. momentum. Using $|V_{cb}|=0.040$ [14],
$\tau(\Lambda_b)=1.07\times 10^{-12}s$ [15], $a_2\sim 0.23$,
\footnote{The parameter $a_2$ is extracted from $B\to \psi K(K^*)$ and $B\to
D(D^*)\pi(\rho)$ decays to be $0.227\pm 0.013$ and $0.23\pm 0.06$
respectively [16].}
and $f_\j=395\,$MeV extracted from the observed $J/\psi\to e^+e^-$ rate,
$\Gamma(\j\to e^+e^-)=(5.27\pm 0.37)$ keV [15],
we find
\be
{\cal B}(\b)=\,1.42\times 10^{-2}|\zeta(\omega)|^2=\,(0.7-1.5)\times 10^{-3}.
\en
When the longitudinal polarization of the $\Lambda$ is measured or
anisotropy in angular distribution is produced in a polarized $\Lambda_b$
decay, it is governed by the asymmetry parameter $\alpha$ given by [13]
\be
\alpha=\,{4m^2_\j{\rm Re}(S^*P_2)+2E^2_\j{\rm Re}(S+D)^*P_1\over 2(|S|^2+
|P_2|^2)m^2_\j+(|S+D|^2+|P_1|^2)E^2_\j}.
\en
Numerically, it reads
\be
\alpha(\b)=\,0.18\,.
\en

    The above prediction for the branching ratio is smaller than our previous
estimate $4.6\times 10^{-3}$ [5] due to the decrease of (i) the baryonic
Isgur-Wise function $\zeta(\omega)$ from 0.53 to 0.23-0.32, (ii) the lifetime
$\tau(\Lambda_b)$ from $1.34\times 10^{-12}s$ to $1.07\times 10^{-12}s$,
and (iii) $|V_{cb}|$ from 0.044 to 0.040. On the other hand, $f_1$ and $g_1$
are increased by 20\% owing to the replacement of $(1-m_\Lambda/m_{\Lambda
_b})$ by $(\omega-m_\Lambda/m_{\Lambda_b})$ [see the footnote \#2 right before
Eq.(10)]. The net result is a decrease of the branching ratio for $\b$.

     We next turn to the second method in which only the $b$ quark is treated
as heavy. Since this method is already elucidated in Ref.[7], we will
recapitulate it here. In the heavy $b$-qaurk limit,
there are only two independent form factors $F_1$ and $F_2$ [17]:
\be
\la\Lambda(p)|\bar{s}\gamma_\mu(1-\gamma_5)b|\Lambda_b(v)\ra=\,\bar{u}
_{_\Lambda}\left(F_1^{\Lambda_b\Lambda}(v\cdot p)+v\!\!\!/ F_2^{\Lambda_b
\Lambda}(v\cdot p)\right)\gamma_\mu(1-\gamma_5)u_{_{\Lambda_b}},
\en
which are related to the standard form factors by
\be
&& f_1(q^2)=g_1(q^2)=F_1(q^2)+{m_\Lambda\over m_{\Lambda_b}}F_2(q^2), \non \\
&& f_2(q^2)=g_2(q^2)=f_3(q^2)=g_3(q^2)=\,{1\over m_{\Lambda_b}}F_2(q^2).
\en
Form factors $f_i$ and $g_i$ can be calculated in the quark model (e.g.
the non-relativistic quark model and the MIT bag model) by first evaluating
them at zero recoil, where the use of the quark model is supposed to be
most trustworthy, and then
extrapolating them to the desired $q^2$ under some assumption on their $q^2$
dependence [18]. We now try to reduce the model dependence by considering
the form-factor ratio $R\equiv F_2/F_1$. Note that the same ratio for
$\Lambda_c\to\Lambda$ transition has been measured recently by CLEO to be [19]
\be
R_{\Lambda_c\Lambda}=-0.33\pm 0.16\pm 0.15\,.
\en
In the heavy quark limit, $R_{\Lambda_b\Lambda}$ is identical to $R_{\Lambda_c
\Lambda}$ at the same heavy quark velocity. At zero recoil or $q^2=q_m^2
\equiv(m_{\Lambda_b}-m_\Lambda)^2$, $f_1$ is reduced to the usual vector
coupling constant, which is found to be 0.95 in the MIT bag model [7].
\footnote{The form factor $g_1$ is essentially reduced to the axial vector
coupling constant. We found $g_1(q_m^2)=0.86$ in the bag model. In the
heavy quark effective theory, the difference between $f_i$ and $g_i$ arises
from $1/m_b$ corrections.}
Substituting this and $R_{\Lambda_b\Lambda}$ into (19) leads to
\be
g_1(q_m^2)=f_1(q_m^2)=0.95\,,~~~g_2(q^2_m)=f_2(q^2_m)=-0.060\,{\rm GeV}^{-1},
\en
where only the central values are used for $f_2$ and $g_2$. The
$q^2$ dependence for form factors is usually parametrized as
\be
f(q^2)=\,f(q_m^2)\left({1-q_m^2/m_*^2\over 1-q^2/m_*^2}\right)^n\equiv
f(q_m^2)\hat{\zeta}(q^2),
\en
where $m_*$ is a pole mass and $\hat{\zeta}(q^2)$ plays the same role as
the Isgur-Wise function $\zeta(v\cdot v')$.
In practice we will take $n=1,2$, corresponding to
monopole and dipole behavior, respectively. Taking $m_*=5.42$ GeV,
\footnote{This is the mass of $B_s^*(1^-)$. The pole mass $m_*$
is the same for $f_i$ and $g_i$ in the heavy quark limit.}
we obtain
\be
{\cal B}(\b) &=& 4.0\times 10^{-3}\left|\hat{\zeta}(m^2_\j)\right|^2 \non \\
&=& \cases{1.6\times 10^{-4} &dipole;  \cr 8.2\times 10^{-4} &monopole, \cr}
\en
and
\be
\alpha(\b)=\,0.31\,.
\en
We see that the branching ratio is very sensitive to the choice of form-factor
$q^2$ dependence.

    To summarize, two different methods have been employed to study the $\b$
decay. The first method is model independent, but it is subject to possible
large higher order $1/m_s$ corrections. The second method takes into
account $m_s$ effects to all orders, but the sensitivity on the
choice of form-factor $q^2$ behavior renders it impossible to make
definite predictions. Nevertheless, we can conclude that ${\cal B}(\b)~{\large
^<_{\sim}}~10^{-3}$ and $\alpha$ is of order $0.20\sim 0.30\,$.

\vskip 1.5 cm
\centerline{\bf ACKNOWLEDGMENT}
     This work was supported in part by the National Science of Council of
ROC under Contract No. NSC84-2112-M-001-014.

\pagebreak
\centerline{\bf REFERENCES}
\vskip 0.3 cm
\begin{enumerate}

\item UA1 Collaboration, C. Albarjar {\it et al.,} \pl {\bf B273}, 540
(1991).

\item CDF Collaboration, F. Abe {\it et al.,} \pr {\bf D47}, 2639 (1993).

\item S.E. Tzmarias, invited talk presented in the 27th International
Conference on High Energy Physics, Glasgow, July 20-27, 1994.

\item See e.g. H.Y. Cheng and B. Tseng, \pr {\bf D46}, 1042 (1992); {\sl ibid}
{\bf D48}, 4188 (1993).

\item H.Y. Cheng, \pl {\bf B289}, 455 (1992).

\item M. Bauer, B. Stech, and M. Wirbel, \zp {\bf C34}, 103 (1987).

\item H.Y. Cheng, C.Y. Cheung, G.L. Lin, Y.C. Lin, T.M. Yan, and H.L. Yu,
CLNS 94/1278 (1994).

\item N. Isgur and M. Wise, \pl {\bf B232}, 113 (1989); {\bf B237}, 527
(1990).

\item H. Georgi, B. Grinstein, and M.B. Wise, \pl {\bf B252}, 456 (1990).

\item A. Falk {\it et al.}, \np {\bf B343}, 1 (1990).

\item E. Jenkins, A. Manohar, and M.B. Wise, \np {\bf B396}, 38 (1993).

\item M. Sadzikowski and K. Zalewski, {\sl Z. Phys.} {\bf C59}, 677 (1993).

\item See e.g. S. Pakvasa, S.F. Tuan, and S.P. Rosen, \pr {\bf D42}, 3746
(1994).

\item M. Neubert, \pl {\bf B338}, 84 (1994).

\item Particle Data Group, \pr {\bf D50}, 1173 (1994).

\item H.Y. Cheng and B. Tseng, IP-ASTP-21-94 (1994).

\item T. Mannel, W. Roberts, and Z. Ryzak, \np {\bf B355}, 38 (1991); \pl
{\bf B255}, 593 (1991); F. Hussain, J.G. K\"orner, M. Kr\"amer, and G.
Thompson, \zp {\bf C51}, 321 (1991).

\item R. P\'erez-Marcial, R. Huerta, A. Garcia, and M. Avila-Aoki, \pr
{\bf D40}, 2955 (1989).

\item CLEO Collaboration, J. Dominick {\it et al.,} CLEO CONF 94-19 (1994).

\end{enumerate}

\end{document}